# Thermal extraction: enhancing thermal emission of finite size macroscopic blackbody to far-field vacuum

Zongfu Yu[1], Nicholas Sergeant[1], Torbjorn Skauli[1,2], Gang Zhang[3], Hailiang Wang[4], and Shanhui Fan[1]

1. Department of Electrical Engineering and Ginzton Lab, Stanford University, Stanford, CA 94305, U.S.A.
2. Norwegian Defense Research Establishment, Norway
3. Department of Electronics, Peking University, Beijing, China
4. Department of Chemistry, Stanford University, Stanford, CA 94305, U.S.A

**The understanding of far-field thermal radiation had directly led to the discovery of quantum mechanics a century ago, and is of great current practical importance for applications in energy conversions, radiative cooling, and thermal control. It is commonly assumed that for any macroscopic thermal emitter, its maximal emitted power within any given frequency range cannot exceed that of a blackbody with the same surface area. In contrast to such conventional wisdom, here we propose, and experimentally demonstrate, that the emitted power from a finite size macroscopic blackbody to far field vacuum can be significantly enhanced, within the constraint of the second law of thermodynamics. To achieve such an enhancement, the thermal body needs to have internal electromagnetic density of states (DOS) greater than that of vacuum, and one needs to provide a thermal extraction mechanism to enable the contributions of all internal modes to far field radiation.**

There has been significant recent development aiming to tailor far-field thermal emission[1]. It has been demonstrated that, for macroscopic emitters that have at least one of the dimensions exceeding several wavelengths, both the spatial directivity and the spectrum of far-field thermal emission can be greatly modified with structures such as photonic crystals, cavity arrays, and metamaterials[1-17]. However, none of these structures can emit more thermal radiation power than a blackbody with the same area[8,18-19]. A blackbody, by definition, has a total thermal emission power $P = \sigma T^4 S$ to far field vacuum, where $S$ is the area of the blackbody, a result that is commonly referred to as the Stefan-Boltzman law. Thus, it is commonly believed that a macroscopic body cannot have its thermal emission power exceed that given by the Stefan-Boltzman law.

The aim of our paper here is to show that a macroscopic blackbody in fact can emit more thermal radiation to far field vacuum than $P = \sigma T^4 S$. The key is to have a macroscopic thermal body with an internal density of states that is higher than vacuum, and to provide a thermal extraction medium that facilitates the out-coupling of these internal states. Importantly, the thermal extraction medium itself is transparent and does not by itself emit any radiation.

Our aim is connected to, but distinct from, two recent developments. In the first development, experiments now have demonstrated that two thermal bodies in close proximity to each other can have thermal conductance exceeding the prediction from the Stefan-Boltzman law[20-24]. Such an enhancement, however, is a purely near-field effect, whereas our focus here is on far-field enhancement. In the second development, it is known that the absorption cross-section of a single optical antenna can significantly exceed its geometric cross-section[18,25-27]. Within a narrow frequency range, the power spectral density of such an emitter can significantly exceed that of a blackbody, if one were to compare the emitter to a blackbody with the same geometric cross-section[1]. In this situation, the size of the object is typically comparable to or smaller than the wavelengths of the thermal radiation[25]. In contrast to these works on individual thermal antennas, our work here concerns broad-band enhancement for macroscopic finite bodies over the entire thermal wavelength range.

To illustrate our concepts we start by considering a thermal emitter consisting of a small opening on a cavity shown in Fig. 1a. The cavity has vacuum immediately outside. The opening has an area $S$. The inner sidewall of the cavity is made of diffusive reflector that also absorbs light. The opening area is completely dark with emissivity of unity: any light entering through the open area bounces many times and eventually gets absorbed by the sidewall. The cavity is maintained at temperature $T$.

We further assume the cavity is filled with transparent dielectric medium that has a refractive index $n_i$. (Below, we refer the cavity with $n_i = 1$ as an "empty cavity", and with $n_i > 1$ as a "filled cavity".) Perfect anti-reflection is assumed at all interfaces. In this case, it is well known[28] that independent of the value of refractive index $n_i$, the cavity always has the same external thermal emission characteristics with the same total emitted power $\sigma T^4 S$ to far field vacuum. Inside the cavity, the internal thermal radiation energy is dependent on the electromagnetic density of states, which scales as $n_i^3$. However, when $n_i > 1$, total internal reflection, occurring at the interface between the medium inside the cavity and vacuum outside, prevents significant portion of the internal electromagnetic modes from coupling to vacuum. The resulting thermal emission to the far field thus has the same profile independent of internal radiation density.

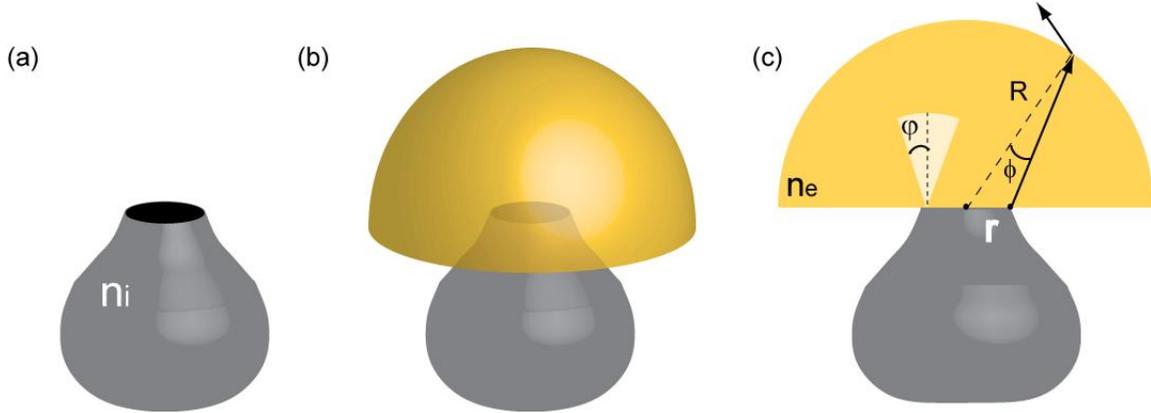

Fig. 1 a) Emitter formed by an open area (black surface) of an absorptive cavity. The cavity can be filled with transparent dielectric of refractive index $n_i$. b) Thermal extraction using a hemispherical dome placed at the opening of the cavity. The dome is transparent and does not emit or absorb any thermal radiations, and has a refractive index $n_e$. c) Cross section showing emission cone (white) of the thermal radiation inside the dome.

As the main result of our paper, we now show and demonstrate that for the emitter in Fig. 1a, one can enhance its thermal emission by placing a hemispherical dome with refractive index $n_e$ covering the entrance of the cavity, as shown in Fig. 1(b). The dome is in close contact with the open area of the cavity. Here the dome plays the role of a thermal extraction device that enables all modes inside the cavity to escape into vacuum. Importantly, the dome itself is assumed to be transparent, so that it does not emit or absorb any thermal radiation.

To calculate the thermal emission from the geometry shown in Fig. 1b, we follow a ray tracing procedure. For simplicity, we assume that the opening of the cavity has a circular shape of radius $r$ and the dome has a radius $R$. We further assume $R \geq n_e r$, which is sufficient to ensure that any light ray originated from the open area $S$, when it reaches the top surface of the dome, has an incident angle less than the total internal reflection angle $\phi \leq \sin^{-1}(1/n_e)$ (Fig. 1c, solid arrows), and therefore can escape to far field vacuum (See Supplementary Information). Here again we assume perfect anti-reflection at the dome surface.

The emission from the cavity forms a light cone in the dome. Half apex angle of the light cone is given by

$$\varphi = \begin{cases} \sin^{-1}(n_i / n_e) & \text{if } n_i \leq n_e \\ \pi/2 & \text{if } n_i > n_e \end{cases} \qquad \text{Eq. (1)}$$

To obtain the total emission power, we integrate thermal radiation within the light cone,

$$P = \begin{cases} n_e^2 \sigma T^4 \pi r^2 & (n_e < n_i) \\ n_e^2 \sigma T^4 \pi r^2 \dfrac{\int_0^{\varphi_2} \sin(\theta)\cos(\theta)d\theta}{\int_0^{\pi/2} \sin(\theta)\cos(\theta)d\theta} = n_i^2 \sigma T^4 \pi r^2 & (n_e \geq n_i) \end{cases} \quad \text{Eq. (2)}$$

Since all lights in the dome can escape without total internal reflection, Eq. (2) is the total thermal radiation to far field vacuum.

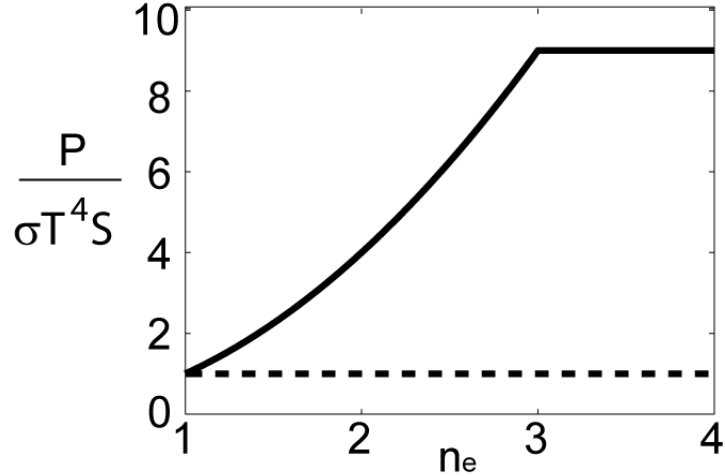

Fig. 2 Total thermal radiation power to far-field vacuum for the structure shown in Fig. 1b, as a function of the refractive index of the dome $n_e$ Solid line: Filled cavity with $n_i = 3$. Dashed line: Empty cavity with $n_i = 1$.

Fig. 2 shows the thermal radiation power $P$ as a function of the refractive index of the dome $n_e$. The total radiation power from an empty cavity (dashed line in Fig. 2) does not change as a function of $n_e$, while the power from the filled cavity (solid line in Fig. 2) increases as $n_e$ increases until $n_e = n_i$. With the assistance of the thermal extraction, the filled cavity can emit up to $n_i^2 S \sigma T^4$ to far field vacuum, $n_i^2$ times of the emission of a blackbody of the same area $S$.

It is well known that when in contact with a transparent medium with index higher than vacuum, an ideal blackbody emits more thermal radiation into the transparent medium as compared to the same blackbody to vacuum[25]. Our use of hemispherical dome exploits this fact, and also ensures that all radiation into the high-index dome can escape to vacuum, leading to enhanced thermal radiation to vacuum. Moreover, our theory above indicates that the internal density of state of the blackbody is in fact important to achieve such emission enhancement. As can be seen in Fig. 2, the internal density of state of the blackbody must be higher than that of the transparent medium in order to achieve the maximum effect of enhanced thermal emission.

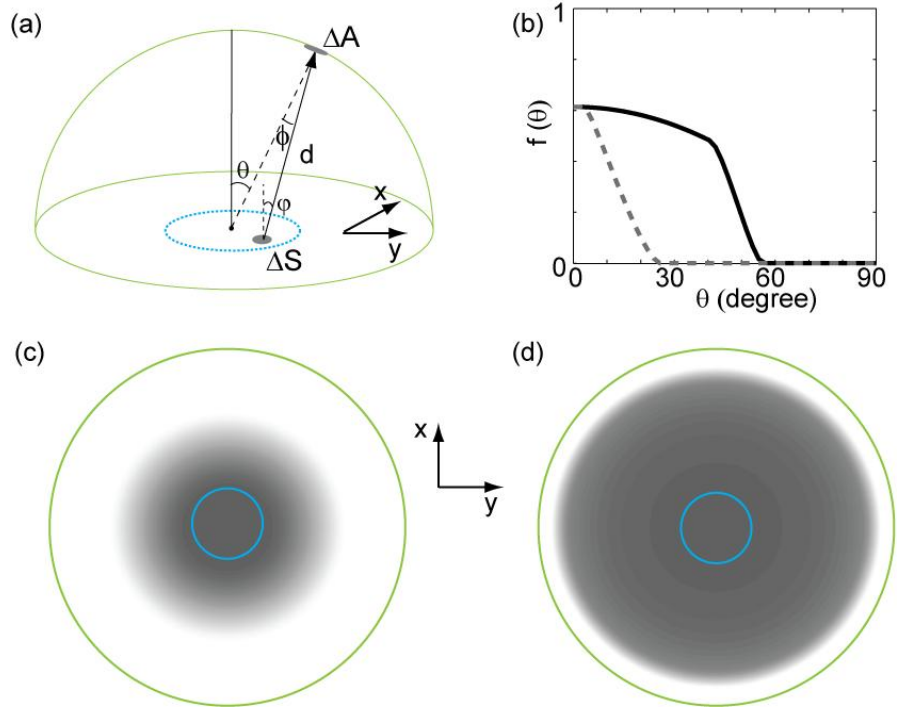

Fig. 3 Distribution of thermal radiation on the surface of the dome for the structure shown in Fig. 1b. a) Schematic of the calculation. Solid arrow indicates radiation that is emitted from a small area $\Delta S$ on the cavity opening and is received by a small area $\Delta A$ on the surface of the dome. b) Distribution of the radiation power on the surface of the dome as a function of the polar angle $\theta$. Dashed and solid lines are for the cases with empty and filled cavities respectively. c,d) The distribution of the radiation power, plotted on the hemisphere of the dome, for the case of empty (c) and filled (d) cavity. The blue circle indicates the emitter area, i.e. the opening area of the cavity. The green circle is the boundary of the dome. Darker region indicates higher emission and the white region has zero emission.

The distribution of thermal radiation on the surface of the dome can be calculated with a schematic shown in Fig. 3a. For a small area $\Delta A$ on the surface of the dome, the thermal radiation power it receives from the cavity is

$$v_{\Delta A} = \Delta A \sigma T^4 \iint_{\varphi < \varphi_1 \text{ and } x^2+y^2 \leq r^2} \cos(\varphi)\cos(\phi) n_e^2 \frac{dxdy}{\pi d^2} \qquad \text{Eq.(3)}$$

(The definitions of the geometric parameters in Eq. (3) are provided in Fig. 3a.) All radiation that the area $\Delta A$ receives can escape the dome. As a dimensionless quantity, we define a normalized power distribution

$$f(\theta) \equiv \frac{v_{\Delta A}}{\Delta A \sigma T^4} \quad , \qquad \text{Eq.(4)}$$

to describe the power distribution on the surface of the dome. Due to rotational symmetry, $f$ depends only on the polar angle $\theta$.

As one specific numerical example, we calculate a case where $n_e = 4$. We choose the dome radius $R = 5r$ to satisfy the condition $R \geq n_e r$. $f(\theta)$ is numerically evaluated using Eq. (3) and is shown in Fig. 3b for both empty cavity ($n_i = 1$), and filled cavity ($n_i = 3$). For both cases, $f(\theta)$ maximizes at normal direction $\theta = 0$, since this area is directly above the opening the cavity. $f(\theta)$ decreases as $\theta$ increases and eventually vanishes for large $\theta$, since at large $\theta$ the corresponding area lies outside the emission cone of cavity. However, the emission profile of the filled cavity expands to a much wider angular range than that of the empty cavity (Fig. 4c, d). Therefore, when thermal extraction occurs, as for example for the case here of the filled cavity, the entire dome appears bright.

Hemispherical dome has been previously used for enhancing output efficiency of light emitting diode[29-30], and as a solid immersion lens for resolution enhancement[31]. However, our main finding, that hemispherical dome can be used to enhance thermal emission, has never been recognized before. The effective emission area of our structure, while greatly exceeding the geometric area of the emitter itself, does not exceed the surface area of the hemispherical dome, as required by the second law of thermodynamics. Our finding here is consistent with known physics of radiometry including the consideration of optical etendue conservation[32].

Based on the theoretical analysis above, we now experimentally demonstrate the use of thermal extraction to enhance thermal emission beyond the blackbody limit. Carbon black paint is used as the thermal emission source. It has a refractive index around 2.3 and an emissivity around 0.85 in the near to mid infrared regime. A circular carbon dot is coated on a polished Al sample holder, which provides a low emission background. The dot has a radius r = 1.025mm with an area of 3.3 mm$^2$ (Fig. 4a). For the thermal extraction device, we use a hemispherical dome made from ZnSe, a transparent material with negligible thermal emission in the near to mid IR region. The hemisphere has a diameter of 6mm and refractive index 2.4 (Fig. 4b). It satisfies the condition $R \geq n_e r$, which ensures that light entering into the hemisphere from the thermal emitter is not trapped by the total internal reflection at the dome interface.

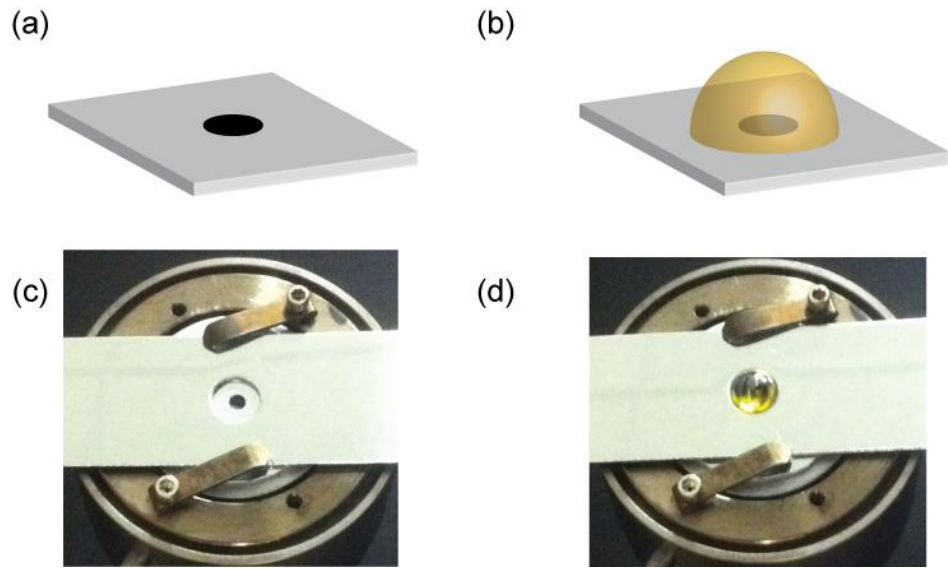

Fig.4 Schematic and actual experimental structure for demonstration of thermal extraction. a,c) Emission source made of carbon dot is coated on an aluminum plate placed on a temperature controlled heater. b,d) Thermal extraction medium made of ZnSe hemisphere is placed in close contact with carbon dot.

The aluminum sample holder is placed on a temperature-controlled heater (Fig. 4c, d). The entire heater is placed in a vacuum chamber (~$10^{-6}$ torr) to avoid oxidation of ZnSe and to maintain thermal stability. The source is observed through a $CaF_2$ window on the vacuum chamber. The thermal emission is collected by a parabolic mirror and sent through an aperture to a Fourier transform infrared (FTIR) spectrometer. The use of the aperture allows us to collect emission from only a small area on the sample holder. For each measurement, we center the collection area on the source by adjusting the position of the aperture until maximum reading is reached. To measure the angular emission, the heater stage can be rotated inside the vacuum chamber. For calibration purposes, a blackbody simulator (Infrared System Development Corporation 564/301 and IR-301 Blackbody controller) is also measured using the same optical setup. By comparing to such calibration measurement, we can therefore obtain the absolute power emission from our thermal emitters (Supplementary Information). The sample is maintained at 553K. Temperature consistency is confirmed by both the thermal controller and the emission spectra (Supplementary Information). Background emission from the Al sample holder is also characterized and has been subtracted out in the data shown below (Supplementary Information).

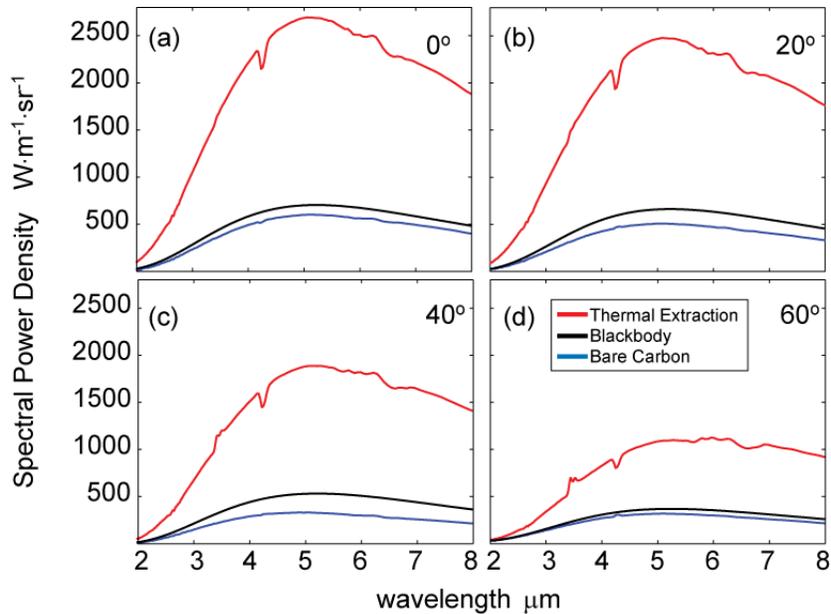

Fig. 5 Emitted power spectra measured at 553K for collection angles of 0, 20, 40 and 60 degrees. Red and blue lines are for the carbon dot with and without the hemispherical dome, respectively. Black lines are emission power from an ideal blackbody of the same area at the same temperature. The ripples in the curves are caused by atmosphere absorption.

Figure 5 shows the emission spectra of the structures at various angles. The emission peaks at 5.25 $\mu m$ wavelength, as expected for a near-black emitter at a temperature of 553K. The spectral density from an ideal blackbody of the same size is plotted as reference (black lines). As expected, the bare carbon dot (blue lines) emits less than the ideal blackbody, with an emissivity of 0.85 in the normal direction. In the presence of the dome, the emitted power in the normal direction from the same carbon black dot is enhanced by 4.46 fold, representing a 3.79 fold enhancement over the emission by an ideal blackbody (Fig. 5a). Similar enhancement is observed for off normal directions as well (Fig. 5b,c,d). The enhanced emission is purely from the extraction of carbon's internal thermal energy, not from the ZnSe hemisphere. To verify this, a reference sample with the ZnSe hemisphere but without the carbon dot is measured, and it shows negligible emission (Supplementary Information).

Figure 6 shows the angular emission, as obtained by integrating the spectral density over the wavelength range of 2 to 8 $\mu m$. For all angles, the presence of the dome results in enhanced emission (Fig. 6, red curve), as compared to both the carbon black dot without the dome (Fig. 6, blue curve) and an ideal blackbody (Fig. 6, black curve) of the same area. The total thermal emission is obtained by integrating over all angles and all wavelengths in the range of 2 to 8 $\mu m$. The total emission is 10.4mW for the ideal black body, 7.6mW for the bare carbon dot, and 31.3mW for the carbon dot with the dome. We have therefore unambiguously demonstrated that one can indeed significantly enhance the emission of a thermal body beyond the blackbody emission with the thermal extraction scheme.

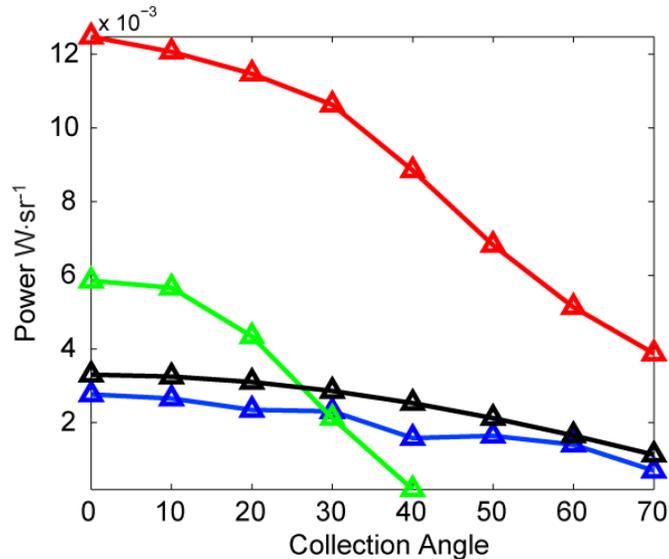

Fig. 6 Experimentally measured emission power from the carbon dot as a function of angle, with the ZnSe hemispherical dome in optical contact (red), without the dome (blue), and with the flat surface of the dome separated from the carbon dot by 30 micron (green). Triangles are measured data points. The black line is the emission power from an ideal blackbody with the same area.

The hemispherical dome we use here is a focusing lens. However, the thermal extraction effect is fundamentally different from the focusing effect of the lens. To achieve thermal extraction, we require that all internal states of the emitters can couple into the modes inside the dome. Thus the emitter and the dome must be in optical contact, i.e. the distance between the emitter and the flat surface of the dome must be significantly smaller than the thermal wavelength. Preventing the optical contact between the emitter and the dome should eliminate the thermal extraction effect. As a demonstration, we conduct a comparison experiment where the dome is lifted away from the carbon dot by 30 $\mu m$, a distance that is large enough to prevent photon tunneling between the emitter and the dome, and small enough to preserve all other geometrical optical lens effects (Fig. 7c). The resulting emission power is shown as green line in Fig. 6. In the normal direction, due to the focusing effect of the hemispherical dome, the emission is higher than that that of bare carbon dot, but it quickly diminishes at large angles with negligible emission beyond 40 degree. Therefore, the geometrical optical effect of ZnSe hemispherical dome can only redistribute the thermal emission but does not enhance the total emission. The total emitted power is only 4.1mW, below that from the blackbody with the same area at the same temperature.

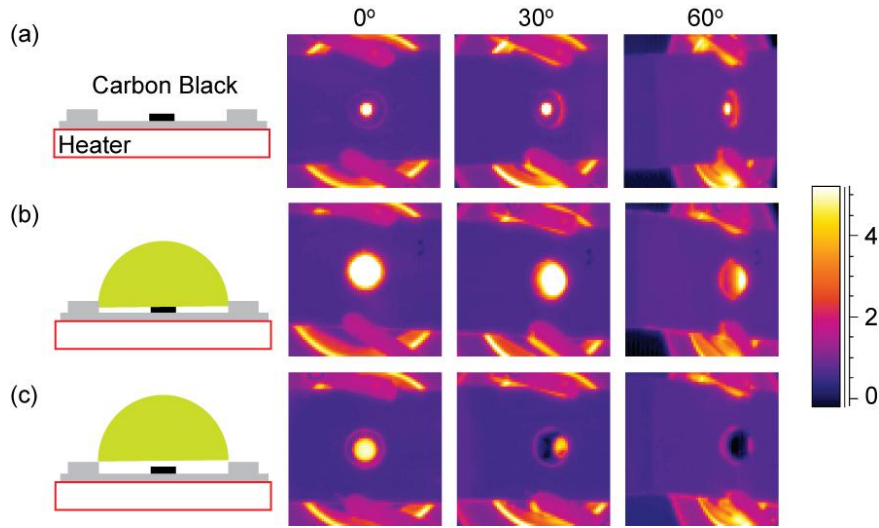

Fig. 7 Infrared images of the thermal sources maintained at a temperature of 553K. Images are taken at 0, 30 and 60 degrees. All images have the same color scale. Values on the color scale bar are linearly proportional to the photon counts of detectors in the camera. In all images, the most outer bright regions are the heater surface beneath the Al sample holder. a) Bare carbon dot. The Al plate has a holder for the hemisphere, the edge of which is visible due to its slightly higher emissivity. b) The carbon dot is in optical contact with ZnSe. Notice that the entire dome lights up at the normal direction and significant emission even at 60 degree angle, demonstrating thermal extraction. c) ZnSe hemisphere is spaced away from carbon dot by 30 $\mu m$, a distance sufficient to prevent thermal extraction. In this case, ZnSe hemisphere only redistributes emission among different directions. More images available in Supplementary Information.

As we have seen in the analysis of Fig. 3, the thermal extraction effect is directly correlated with a broadening of angular distribution of photons on the dome surface. When thermal extraction occurs, the apparent emitting area thus should appear larger *from all viewing angles*. To demonstrate this, we directly visualize the emitters with an IR camera (FLIR System Inc. SC4000, spectral range 3 to 5 $\mu m$) (Fig. 7). We compare three cases: the bare carbon dot, the carbon dot in optical contact with the ZnSe dome, and the carbon dot separated from the ZnSe dome by 30 $\mu m$. For all three cases, the emitting sources are the same, but as we have already seen in Figs. 5 and 6, the emitted powers are drastically different. Such differences can be directly visualized with the camera.

The bare carbon dot has an emission profile that is approximately Lambertian. The apparent emitting area reduces with increasing angle (Fig. 7a). When the carbon dot is in optical contact with the hemispherical lens (Fig. 7b), the effect of thermal extraction produces much larger apparent emitting area for all viewing angles. This agrees with the angular distribution calculation shown in Fig. 3.

Fig. 7c shows the case where the ZnSe hemisphere is spaced away from the carbon dot. At normal direction, the apparent emitting area is larger as compared with that of the bare dot. But the apparent emitting area decreases very rapidly with angles. At large angles, the apparent emitter area is smaller compared with the case of the bare dot. This again is consistent with the results of Fig. 6, showing that thermal extraction is fundamentally different from a focusing effect.

We have illustrated the concept of thermal extraction with the example of a high-index hemispherical dome. In general, thermal extraction can be accomplished with other geometries as well. Here we comment on the general requirement of the thermal extraction medium:

Firstly, the thermal extraction needs to be in optical contact with the emitter, i.e. the distance between the emitter and the extraction medium needs to be smaller than the thermal wavelength $\lambda_T = \hbar c / k_B T$. This is to ensure that all internal states in the emitter can couple to modes in the extraction medium. We note, however, the thermal extraction medium needs not be in physical contact with the emitter. This could be useful in practice when it is advantageous to prevent thermal conduction between the extraction medium and the emitter.

Secondly, from a thermodynamics point of view, the thermal extraction medium needs to provide enough radiation channels [33] over the area of the emitter to ensure that all internal modes of the emitter can out-couple. A simple way to accomplish this is to choose the extraction medium such that its density of states is larger than that of the emitter. The size of the extraction medium also needs to be sufficiently large, such that the vacuum region immediately outside the extraction medium has sufficient number of radiation channels to accommodate all the thermal emission. Both of these considerations are incorporated in our choice of parameters for the hemispherical dome. On the other hand, based upon these considerations, one can envision a wide variety of nanophotonic structures that may satisfy these thermodynamic considerations. For example, structured materials, e.g. photonic crystal and metamaterial, can be engineered to have extremely high channel density. Their dispersion relations can also be tailored such that extracted radiations can be guided to vacuum interfaces where enough channels in vacuum are available to accommodate the radiation.

Finally, in the extraction medium those optical modes that receive radiation from the emitter need to be accessible to far field vacuum. This places a constraint on the geometry of the extraction medium. For example, a transparent high index slab with flat surface does not provide thermal extraction. Even though more radiation can enter the slab, those outside the escape cone defined by $\sin^{-1}(1/n_e)$ cannot escape to far field vacuum due to total internal reflection. As a result, the total far field emission remains the same as $S\sigma T^4$. This particular requirement on making internal optical states accessible to far field shares the same spirit of the requirement on light trapping in solar cells [33-35]. Many light trapping structures, for example, roughened slab, irregular polygon, and nanostructured interface can be directly used for thermal extraction.

The demonstration of thermal extraction here opens possibilities for a number of applications. For example, there is a strong effort seeking to miniaturize the active emitting region of a thermal source, since with a smaller active region[36], it takes less power to drive the active region to a prescribed temperature. However, at a constant temperature miniaturization of the active region typically comes with the price of reduction in emitted power. Here we show that it is actually possible to decouple the area of the active emitter and its emitted power, which may potentially lead to a better thermal emitter with higher power efficiency. Thermal extraction also indicates the possibility of enhancing thermal transport in the far field for efficient radiative cooling and heating[37].

This work is supported in part by the Global Climate and Energy Project (GCEP) of the Stanford University.

References:

1   Greffet, J.-J. Applied physics: Controlled incandescence. *Nature* **478**, 191-192 (2011).
2   Cornelius, C. M. & Dowling, J. P. Modification of Planck blackbody radiation by photonic band-gap structures. *Phys. Rev. A* **59**, 4736-4746 (1999).
3   Maruyama, S., Kashiwa, T., Yugami, H. & Esashi, M. Thermal radiation from two-dimensionally confined modes in microcavities. *App. Phys. Lett.* **79**, 1393-1395 (2001).
4   Greffet, J.-J. *et al.* Coherent emission of light by thermal sources. *Nature* **416**, 61-64 (2002).
5   Pralle, M. U. *et al.* Photonic crystal enhanced narrow-band infrared emitters. *App. Phys. Lett.* **81**, 4685-4687 (2002).
6   Fleming, J. G., Lin, S. Y., El-Kady, I., Biswas, R. & Ho, K. M. All-metallic three-dimensional photonic crystals with a large infrared bandgap. *Nature* **417**, 52-55 (2002).
7   Lin, S.-Y., Fleming, J. G. & El-Kady, I. Three-dimensional photonic-crystal emission through thermal excitation. *Opt. Lett.* **28**, 1909-1911 (2003).
8   Luo, C., Narayanaswamy, A., Chen, G. & Joannopoulos, J. D. Thermal Radiation from Photonic Crystals: A Direct Calculation. *Phys. Rev. Lett.* **93**, 213905 (2004).
9   Lee, B. J., Fu, C. J. & Zhang, Z. M. Coherent thermal emission from one-dimensional photonic crystals. *App. Phys. Lett* **87**, 071904-071903 (2005).
10  Celanovic, I., Perreault, D. & Kassakian, J. Resonant-cavity enhanced thermal emission. *Phys. Rev. B* **72**, 075127 (2005).
11  Laroche, M. *et al.* Highly directional radiation generated by a tungsten thermal source. *Opt. Lett.* **30**, 2623-2625 (2005).
12  Dahan, N. *et al.* Enhanced coherency of thermal emission: Beyond the limitation imposed by delocalized surface waves. *Phys. Rev. B* **76**, 045427 (2007).
13  Wang, C.-M. *et al.* Reflection and emission properties of an infraredemitter. *Opt. Express* **15**, 14673-14678 (2007).
14  Puscasu, I. & Schaich, W. L. Narrow-band, tunable infrared emission from arrays of microstrip patches. *Appl. Phys. Lett.* **92**, 233102 (2008).

15  Rephaeli, E. & Fan, S. Absorber and emitter for solar thermo-photovoltaic systems to achieve efficiency exceeding the Shockley-Queisser limit. *Opt. Express* **17**, 15145-15159 (2009).
16  Liu, X. *et al.* Taming the Blackbody with Infrared Metamaterials as Selective Thermal Emitters. *Phys. Rev. Lett.* **107**, 045901 (2011).
17  Yeng, Y. X. *et al.* Enabling high-temperature nanophotonics for energy applications. *Proc. Natl. Acad. Sci.* 1120149109 (2012).
18  Trupke, T., Wurfel, P. & Green, M. A. Comment on ``Three-dimensional photonic-crystal emitter for thermal photovoltaic power generation'' *Appl. Phys. Lett.* **84**, 1997-1998 (2004).
19  Fleming, J. G. Addendum: ``Three-Dimensional Photonic-Crystal Emitter For Thermal Photovoltaic Power Generation''  *Appl. Phys. Lett.* **86**, 249902 (2005).
20  Polder, D. & Van Hove, M. Theory of Radiative Heat Transfer between Closely Spaced Bodies. *Phys. Rev. B* **4**, 3303-3314 (1971).
21  Pendry, J. Radiative Exchange of Heat Between Nanostructures.  J. Phys.: Condens. Matter. **11** 6621-6633(1999).
22  Volokitin, A. I. & Persson, B. N. J. Near-field radiative heat transfer and noncontact friction. *Rev. Mod. Phys.* **79**, 1291-1329 (2007).
23  Rousseau, E. *et al.* Radiative heat transfer at the nanoscale. *Nat Photon* **3**, 514-517, (2009).
24  Shen, S., Narayanaswamy, A. & Chen, G. Surface Phonon Polaritons Mediated Energy Transfer between Nanoscale Gaps. *Nano Letters* **9**, 2909-2913, (2009).
25  Zhang, Z. *Nano Microscale Heat Transfer*.  305 (McGraw-Hill, 2007).
26  Ingvarsson, S., Klein, L., Au, Y.-Y., Lacey, J. A. & Hamann, H. F. Enhanced thermal emission from individual antenna-like nanoheaters. *Opt. Express* **15**, 11249-11254 (2007).
27  Schuller, J. A., Taubner, T. & Brongersma, M. L. Optical antenna thermal emitters. *Nat Photon* **3**, 658-661, (2009).
28  Sigel, R. & Howell, J. R. *Thermal Radiation Heat Transfer  (3rd)*. 3 edn,  37 (Hemisphere Publishing Corporation).
29  Carr, W. N. & Pittman, G. E. One-watt GaAs p-n junction infrared source. *Appl. Phys. Lett.* **3**, 173-175 (1963).
30  Moreno, I., Bermúdez, D. & Avendaño-Alejo, M. Light-emitting diode spherical packages: an equation for the light transmission efficiency. *Appl. Opt.* **49**, 12-20 (2010).
31  Terris, B. D., Mamin, H. J., Rugar, D., Studenmund, W. R. & Kino, G. S. Near-field optical data storage using a solid immersion lens. *Appl. Phys. Lett.* **65**, 388-390 (1994).
32  Winston, R., Minano, J. C. & Benitez, P. *Nonimaging Optics*.  pp18 (Academic Press, 2004).
33  Yu, Z., Raman, A. & Fan, S. Fundamental limit of nanophotonic light trapping for solar cells. *Proc. Nat. Acad. Sci. USA* **107**, 17491-17496 (2010).
34  Yablonovitch, E. Statistical ray optics. *J. Opt. Soc. Am. A* **72**, 899-907 (1982).
35  Campbell, P. & Green, M. A. Light trapping properties of pyramidally textured surfaces. *J. Appl. Phys.* **62**, 243-249 (1987).

36  De Zoysa, M. *et al.* Conversion of broadband to narrowband thermal emission through energy recycling. *Nat Photon* **6**, 535-539 (2012).
37  Catalanotti, S. *et al.* The radiative cooling of selective surfaces. *Solar Energy* **17**, 83-89 (1975).

# Supplementary Information

Thermal extraction: enhancing thermal emission of finite size macroscopic blackbody to far-field vacuum

Zongfu Yu[1], Nicholas Sergeant[1], Torbjorn Skauli[1,2], Gang Zhang[3], Hailiang Wang[4], and Shanhui Fan[1]

1. Department of Electrical Engineering and Ginzton Lab, Stanford University, Stanford, CA 94305, U.S.A.
2. Norwegian Defense Research Establishment, Norway
3. Department of Electronics, Peking University, Beijing, China
4. Department of Chemistry, Stanford University, Stanford, CA 94305, U.S.A

**Derivation of the total internal reflection condition in a hemispherical dome**

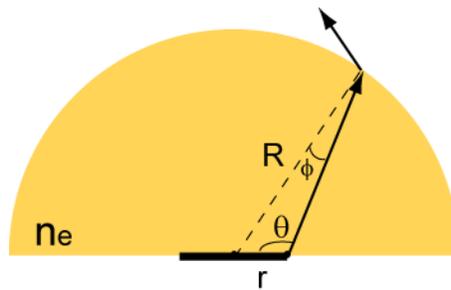

Fig. S1. Schematic showing that light originated from the source at the bottom of a hemispherical dome (thick black line) escapes the dome without total internal reflection.

In this section, we consider the geometry shown in Fig. S1, and derive the condition that allows emission from the source area to escape from the dome without total internal reflection. The emitting area has a radius $r$. The dome has a radius $R$. At the surface of the dome, the incident angle of the ray is $\phi$. Using the sine rule, we have

$$\sin(\phi) = \frac{r \sin(\theta)}{R} \qquad \text{Eq. (S1)}$$

As a worst-case scenario, we assume a maximum emission cone from the source to the dome, i.e. the maximum value for $\theta$ is $90°$. To prevent the total internal reflection, we require that $\sin(\phi) \leq 1/n_e$. With these considerations, and using Eq. S1, we therefore have $R \geq r n_e$.

**Infrared spectroradiometer**

The measurement of spectral radiance from the heated sample is performed with a custom built vacuum emissometer (Fig. S2a). The sample holder is based on a 1'' diameter high temperature heater (HeatWave Labs Model 104863-02) in an ultrahigh vacuum chamber (Fig. S2b). Samples are mounted on the heater. A feedthrough mechanism allows rotation of the heater in vacuum, in order to measure the angular dependence of the spectral radiance. The axis of rotation is in the plane on which the sample is mounted, thus minimizing sample displacement during rotation. The directional spectral radiance emitted by samples is observed outside the vacuum chamber through a $CaF_2$ window.

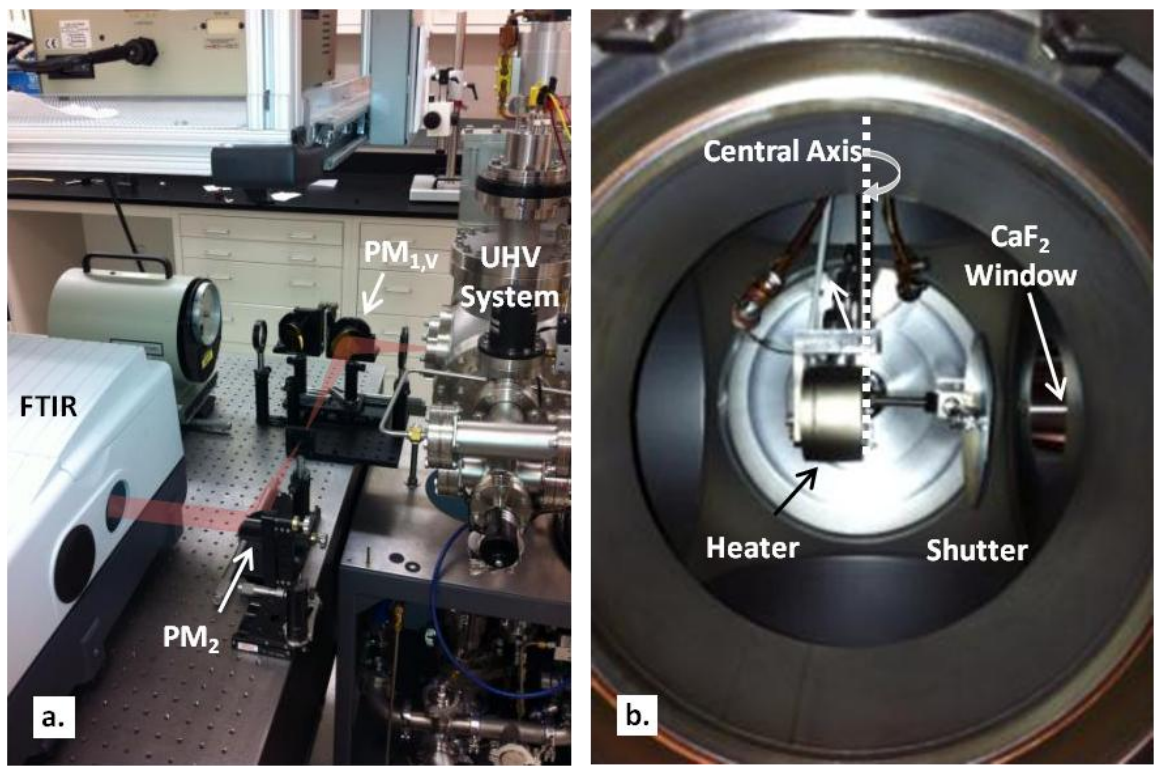

Fig. S2 a) External imaging and collimating optics towards entry port of FTIR spectrometer. b) Inside view of the vacuum emissometer.

External imaging and collimating optics are used to guide the spectral radiance into the entry port of FTIR spectrometer (Fig. S2a) (FTIR: Nicolet 6700 with $CaF_2$ beam splitter and Deuterated Triglycine Sulfate (DTGS) detector with KBr window). The external optical setup consists of a collection and a collimation system. A gold coated 90° off-axis parabolic mirror with focal length $f = 152.4$mm and diameter $D = 50.88$mm (Edmund Optics NT47-110) are used to collect the emitted radiation. This mirror images the sample plane onto a tunable aperture. The aperture is tuned in both diameter and position to select the size and the location of the collection area on the sample. A second gold coated 90° off-axis parabolic mirror with focal length $f = 203.2$mm and diameter $D =$

38.1 mm (Newport 50332AU) is used to collimate the light from the aperture into the entry port of the FTIR spectrometer.

**Transfer function of the measurement system**

The raw data measured by the FTIR system is in an arbitrary linear unit. Fig. S3 shows the raw data measured for the bare carbon dot, and the carbon dot in optical contact with the ZnSe dome, both at a temperature of 553K. To convert this arbitrary unit to power measurement, we need to know the transfer function of the system.

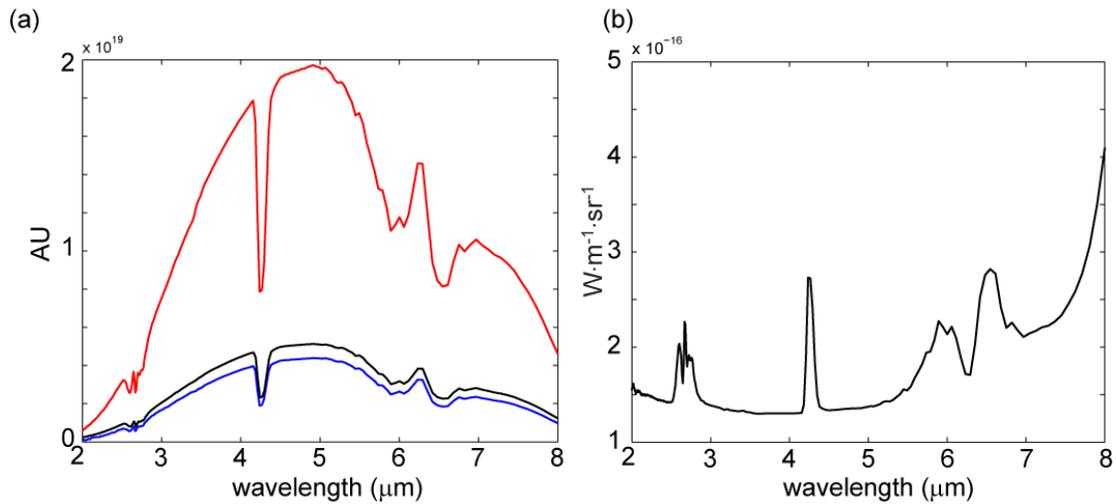

Fig. S3 a) Raw emission spectrum measured from FTIR, for the bare carbon dot (blue), the carbon dot in optical contact with the ZnSe dome (red), and the blackbody simulator (black). The emission areas are 3.3 mm$^2$. Significant atmosphere absorption can be seen in the spectra. b) Transfer function of the system. The spectral densities shown in the main text are obtained by multiplying this transfer function with the raw signal measured from FTIR.

We obtain the transfer function by calibrating the system to a known blackbody emitter simulator (Infrared System Development Corporation 564/301 and IR-301 Blackbody controller). The blackbody simulator consists of a cone cavity with black absorptive inner surface. The open area of the cone has emissivity over 0.99. The emission is collected by the same optical configuration. The collection area is controlled by the aperture size. By assuming the known emitter as an ideal blackbody and thus having a standard Planck's law emission, we can obtain the transfer function of the system shown in Fig. S3b. The transfer function converts the measured FTIR signal to absolute power emission.

**Determination of Temperature**

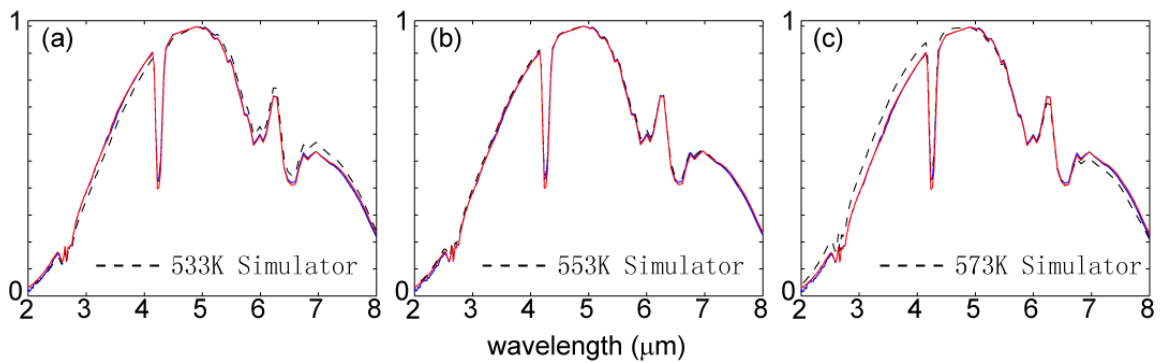

Fig. S4. Normalized spectra measured by FTIR. Red/blue lines are for the bare carbon dot and the carbon dot in optical contact with the dome, respectively, both measured at the same temperature. Dash solid lines in a), b), c) are spectra for the reference blackbody emitter measured at 533K, 553K, and 573K respectively. The 553K has the best match with sample spectra.

The samples are maintained at the same temperature for measurement. The heater controller is set to be the same temperature for all measurement. The samples are placed on the heater surface in vacuum, and thus isolated from heat transfer by conduction and convection. Measurements are performed 45 minutes after the set temperature is reached to help the sample reach steady state.

The temperate consistency for different measurement is confirmed by the lineshapes of the emission. Fig. S4 shows the normalized lineshapes for the emission measured for the bare carbon dot (blue), and the carbon dot in optical contact with the dome (red). These lineshapes overlap very well, showing that they are at the same temperature.

We match the emission lineshapes of the samples to those of a reference blackbody simulator(Infrared System Development Corporation 564/301 ) to obtain the sample surface temperature. This is accurate since the carbon black's emissivity has little wavelength dependency in the measured spectral range. The reference blackbody emitter's temperature is maintained by Infrared System Development Corporation IR-301 Blackbody controller. Spectra of the reference blackbody are measured from 533K to 573K (Fig. S4 dashed lines). The 553K lineshape shows the best match with the sample spectra (Fig. S4a), while 533K and 573K show significant deviation (Fig. S4b, S4c). The sample temperature is determined to be 553 ± 10K.

**Background Emission**

The aluminum sample holder has small but finite emission. In this section, we describe the experiments to determine the background emission by the sample holder including the ZnSe hemisphere. We also show that the ZnSe hemisphere by itself emits very little thermal radiation.

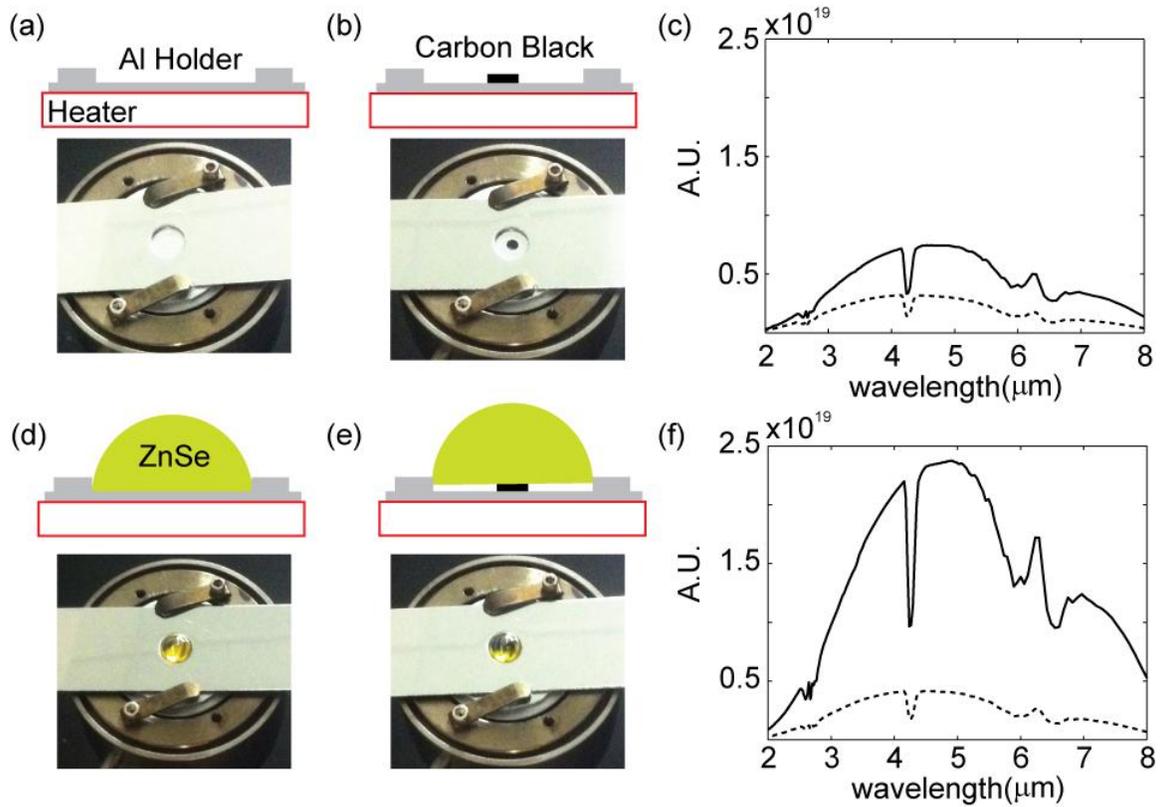

Fig. S5. Experiments to determine the background emission from sample holders and ZnSe hemisphere. (a)-(c): Sample holder without (a) and with (b) the bare carbon dot. c) Emission spectra for the structures in (a) and (b). Dashed and solid lines are for the setup in a) and b) respectively. (d)-(f): Sample with the ZnSe dome, and either without (d) or with (e) the carbon dot. f) Emission spectra of the structures in (d) and (e). Dashed and solid lines are for the setup in d) and e) respectively.

First, the sample holder without carbon dot is measured (Fig. S5a). Its emission is shown by Fig. S5c dashed line. The collection area is slightly larger than the center circular aperture shown in Fig. S5a. The collection area includes a circular edge which is used later on to fix the ZnSe hemisphere. Then, carbon dot is coated on the Al sample holder (Fig. S5b). Despite its small area compared to the collection area, the carbon dot dominates the emission as compared to the background emission from the aluminum sample holder (Fig. S5c solid line). The emission shown in the main text is obtained by subtracting out the background emission. Similarly, measurements are performed for the ZnSe hemisphere without the carbon dot (Fig. S5d). ZnSe does not have significant thermal emission, as can be seen by comparing the dashed lines in Figs. S5c and S5f. The enhanced thermal emission in Fig. S5f, in the presence of the carbon dot, is purely from extracting the internal thermal radiation energy from the carbon dot.

**Infrared Images**

We use an infrared camera based on an InSb photodiode array image sensor with 320x256 pixels. The specified spectral range is 3.0 to 5.0 $\mu m$. Images are read out as raw data, which are proportional to the received number of photons within the spectral range of the camera. Pixel-to-pixel responsivity variations are negligible for the purposes of this work. The camera views the sample through the $CaF_2$ window on the vacuum chamber via a plane gold mirror with high reflectance. A 50 mm focal length f/2.5 lens was used. The sample holder allows observation of the emission source for angles out to about 70 degrees (with some obscuration occurring at larger angles). Images below show the progression of the source emission with increasing off-axis angle. The artificial color scale is the same in all images.

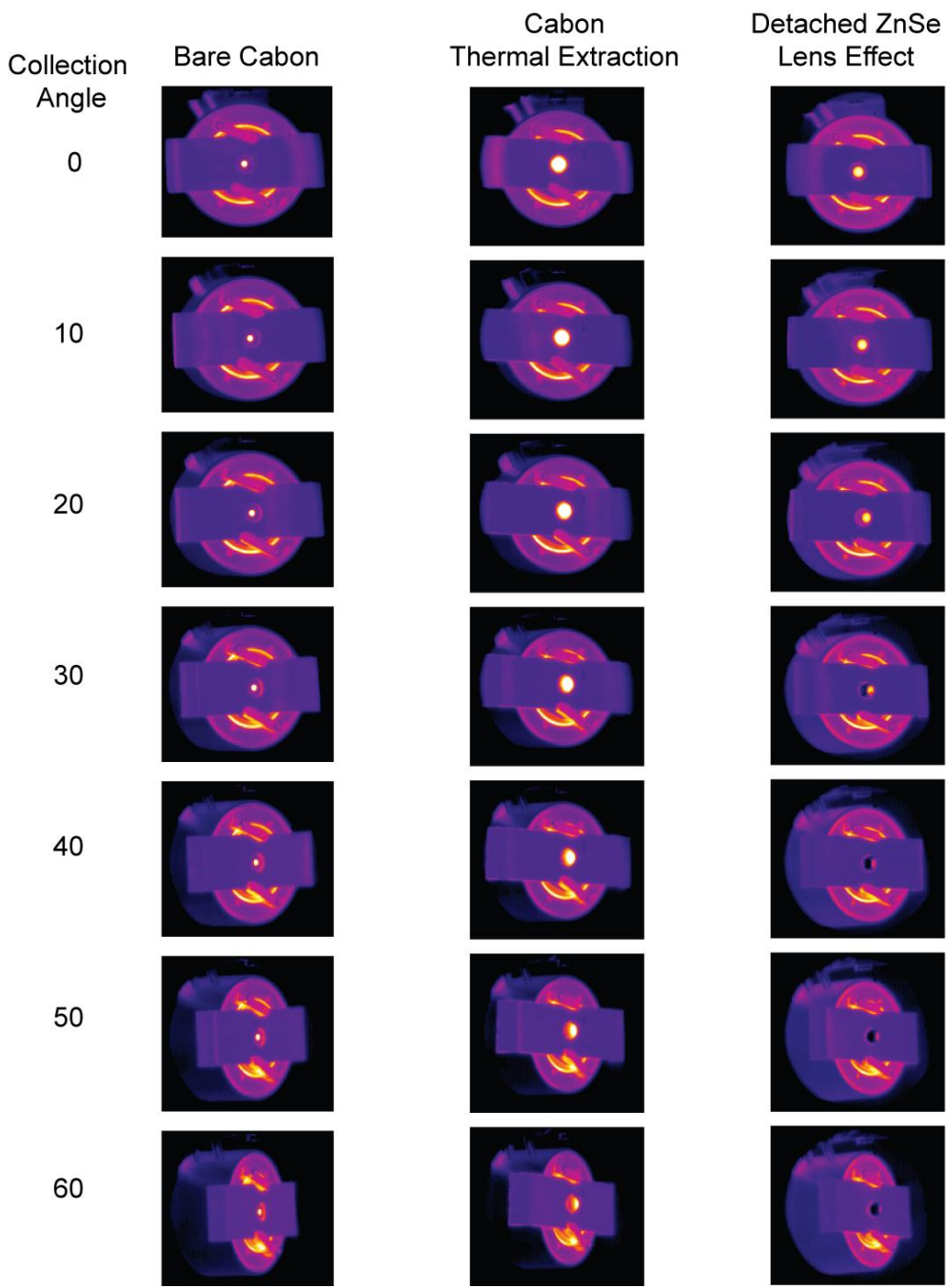

Fig. S6 Infrared camera images for angles from 0 to 60 degrees. The left panels are for a bare carbon dot. The middle panels are for the carbon dot in optical contact with the ZnSe dome. The right panels are for the carbon dot separated from flat surface of the ZnSe dome by 30 μm.